\documentclass[a4paper,english]{aa}
\usepackage{times}
\usepackage[T1]{fontenc}
\usepackage{amsmath}
\usepackage{graphicx}
\usepackage{amssymb}
\usepackage{natbib}
\usepackage{babel}
\usepackage{xspace}
\bibpunct{(}{)}{;}{a}{}{,}
\newcommand{\ascc}   {\mbox{ASCC-2.5}\xspace}
\newcommand{\usnob}   {\mbox{USNO-B1.0}\xspace}

\newcommand{\olin}[1]{\overline{#1}}

\newcommand{\mc}[3]{\multicolumn{#1}{#2}{#3}}
\defcitealias{intpar}{Paper~I}
\defcitealias{clussp}{Paper~II}
\begin{document}

\title{Global survey of star clusters in the Milky Way}

\subtitle{I. The pipeline and fundamental parameters in the second quadrant}

\author{N.V.~Kharchenko \inst{1,2,3} \and
        A.E.~Piskunov \inst{1,2,4} \and
        E.~Schilbach \inst{2} \and
        S.~R\"{o}ser \inst{2} \and
        R.-D.~Scholz \inst{1} }

\offprints{R.-D.~Scholz}

\institute{Leibniz-Institut f\"ur Astrophysik Potsdam (AIP), An der Sternwarte 16, D--14482
Potsdam, Germany\\
email: rdscholz@aip.de
\and
Astronomisches Rechen-Institut, Zentrum f\"ur Astronomie der Universit\"at
Heidelberg, M\"{o}nchhofstra\ss{}e 12-14, D--69120 Heidelberg, Germany
\and
Main Astronomical Observatory, 27 Academica Zabolotnogo Str., 03680 Kiev,
Ukraine
\and
Institute of Astronomy of the Russian Acad. Sci., 48 Pyatnitskaya Str., 109017
Moscow, Russia
}

\date{Received 22 December 2011 / Accepted 7 May 2012}

\abstract{}
{On the basis of the PPMXL star catalogue we performed a survey of star clusters in the second quadrant of the Milky Way.}
{From the PPMXL catalogue of positions and proper motions we took the subset of stars with near-infrared photometry from 2MASS and added the remaining 2MASS stars without proper motions (called 2MAst, i.e. 2MASS with astrometry). We developed a data-processing pipeline including interactive human control of a standardised set of multi-dimensional diagrams to determine kinematic and photometric membership probabilities for stars in a cluster region. The pipeline simultaneously produced the astrophysical parameters of a cluster. From literature we compiled  a target list of presently known open and globular clusters, cluster candidates, associations, and moving groups. From established member stars we derived spatial parameters (coordinates of centres and radii of the main morphological parts of clusters) and cluster kinematics (average proper motions and sometimes radial velocities). For distance, reddening, and age determination we used specific sets of theoretical isochrones. Tidal parameters were obtained by a fit of three-parameter King profiles to the observed density distributions of members.}
{We investigated all 871 objects in the 2nd Galactic quadrant, of which we successfully treated 642 open clusters, 2 globular clusters, and 8 stellar associations. The remaining 219 objects (24\%) were recognised by us to be nonexistent clusters, duplicate entries, or clusters too faint for 2MAst. We found that our sample is complete in the 2nd quadrant up to a distance of 2~kpc, where the average surface density is 94 clusters per kpc$^{2}$. Compared with literature values we found good agreement in spatial and kinematic data, as well as for optical distances and reddening. Small, but systematic offsets were detected in the age determination.}
{}


\maketitle

\section{Introduction}
\label{sec:intro}

Star clusters are important representatives of the galactic population. Their fundamental astrophysical parameters, such as ages and masses, can be determined much more reliably and accurately than for individual stars. In the past, the efforts of astronomers concentrated on the analyses of individual clusters and their properties in our Galaxy and the Magellanic Clouds. During the last few decades the interest has shifted to studies of large samples of clusters, Galactic and extragalactic, which yielded clues for our understanding of the history of star formation and the evolution of galactic populations over large spatial scales. Unlike in our own Galaxy, the distribution of open star clusters in nearby galaxies can be observed over the whole body of a galaxy, which permits direct access to the spatial distribution within the host galaxy. As examples we mention recent studies of various properties of extragalactic clusters in different environments that involved hundreds to a few thousands of objects \citep[see e.g.][]{caldw09, cant09, moraea09, vans09, harrea10, sanrom10, werza11}. The price for this advantage is a poorer resolution (mostly individual stars cannot be resolved), and we have to rely on more sophisticated models such as simple stellar population (SSP) models to estimate fundamental parameters of the clusters.

In the Galaxy, star clusters are resolved into individual stars, and the bulk astrophysical parameters can be derived properly. In the case of open clusters we have to pay for this huge advantage with a limited horizon of a few kiloparsecs around the Sun. In the past, much effort has been devoted to collect different (usually heterogeneous) data on individual clusters to provide a basis for studying their population \citep[][]{beck63,beckfen71,lynga82,lynga5}. The most recent collections of this nature are the updated on-line list of cluster data of \citet{daml02} and the database on Galactic clusters WEBDA\footnote{http://www.univie.ac.at/webda/}. Nowadays, large all-sky photometric and/or kinematic surveys have opened the possibility to derive homogeneous data sets of the parameters of Galactic clusters. Especially the 2MASS survey \citep{cat2MASS} has considerably contributed to the improvement of our knowledge on different aspects of Galactic clusters. Among other studies, we mention the work by \citet{dubi00}, \citet{dubi01}, \citet{dubisb03}, \citet{froeb07}, \citet{froea10}, \citet{gluea10}, \citet{bukea11}, and \citet{tad11}, which deal with large samples of the Galactic star clusters whose determination of the cluster parameters is based on 2MASS data. We also mention studies of smaller samples of clusters \citep{ivbo02,kogl08,mercl05}.

Obviously, the reliability of the results from photometric data can be considerably improved if kinematic information is also included in the membership determination. However, an increasing number of clusters with homogeneous parameters alone is not sufficient for studies of the cluster population: knowing the statistical properties of cluster samples (how complete, how representative) is another basic requirement.

In our previous study we used the \ascc \citep[]{kha01} catalogue to identify known clusters and systematically search for new ones. This resulted in a sample of 650 local open clusters with individual membership information for cluster stars based on kinematic, photometric, and spatial criteria. For each cluster a set of homogeneous parameters was determined including spatial (positions, distances), structural (apparent sizes), kinematic (proper motions and sometimes radial velocities), photometric (reddenings, integrated magnitudes and colours), dynamical (King' tidal parameters and masses) and evolutionary (ages) parameters \citep[][]{clucat,newc109,clumart,clumart1,intpar}. The results are summarised in the Catalogue of Open Cluster Data (COCD), available at the CDS online archive. The cluster sample is complete within about 0.85 kpc and describes the local population of open clusters. However, owing to the bright limiting magnitude of \ascc, this is sufficient only for studying the local inter-arm cluster population of the Galactic disk. To reach larger distances one needs a deeper survey or a shift to less obscured wavelengths.

The current project ``Milky Way Star Clusters'', hereafter referred to as MWSC, is based on the new all-sky catalogue PPMXL \citep[]{ppmxl}, which goes considerably deeper than \ascc, and provides a substantial expansion of the accessible volume. We aim at identifying different cluster-like objects (open and globular clusters and candidates, associations, moving groups, cluster remnants, etc. from a candidate list) in the PPMXL and at determining their fundamental parameters. We use an automated pipeline for membership and cluster parameter determination which is, additionally, under human control at each stage. We expect to considerably increase the number of clusters having a homogeneous set of cluster parameters and reach the nearby spiral arms with the completed part of the survey. The data products will include information on membership of stars in cluster areas (catalogue of stars, MWSCS), and for each confirmed cluster we will provide a set of homogeneous basic cluster parameters (MWSCD), together with an atlas of cluster diagrams (MWSCA). We started at $l=90^\circ$ and are moving along increasing galactic longitude.

The current paper releases the data on all clusters identified within the 2nd Galactic quadrant ($l=90^\circ\dots180^\circ$). In Sect.~\ref{sec:data} we specify the basic input data. The pipeline of the cluster parameter determination is described in detail in Sect.~\ref{sec:pipeline}. In Sect.~\ref{sec:clupop} we estimate the statistical properties of the cluster sample so far and give preliminary results on the spatial ditribution of open clusters in this sector of the Milky Way. Sect.~\ref{sec:conc} summarises the first results of the project.

\section{Data}\label{sec:data}

For our study, the basic data were taken from the all-sky catalogues PPMXL \citep[]{ppmxl} and 2MASS \citep{cat2MASS}. PPMXL gives positions and proper motions in the International Celestial Reference System (ICRS) and low-accuracy photometry from  \usnob \citep[]{usnob1} for about 900~million objects down to $V\approx20$. For some 400~million entries the catalogue  contains accurate $J,H,K_s$ magnitudes from 2MASS \citep{cat2MASS}. We note that the 2MASS limiting magnitudes vary significantly due to crowding. This subset was used to verify the cluster and determine cluster parameters in the astrometric and photometric systems (of PPMXL) that are homogeneous over the whole sky.

\subsection{Construction of 2MAst}

For about 70~million stars from 2MASS no counterparts are given in PPMXL. These stars are either severely reddened along the Galactic disk and bulge, or are extremeley red and too faint in the optical to be detected in \usnob. Another portion of stars is missing in PPMXL because they are blanketed by large images of bright stars with ghosts and spikes appearing on the Schmidt plates. To avoid incompleteness of the survey in such areas, we kept these 2MASS stars  with their 2MASS photometry and used them for photometric selection of cluster members. In case of extended objects, PPMXL and 2MASS (though the latter less often) may include spurious entries and artefacts, and about 12\% of 2MASS objects have more than one \usnob counterpart \citep[]{ppmxl}. This occurs mainly because of the \usnob matching problems in areas where the Schmidt plates of the Palomar survey overlap. In these cases we averaged the corresponding proper motions of the PPMXL. 

The resulting data set 2MAst (stands for 2MASS with astrometric data) includes about 471~million stars. For about 399~million stars it  contains the coordinates and proper motions from  PPMXL, and near infra-red (NIR) photometry and flags from 2MASS. For the remaining stars, only the original information from 2MASS is available. The limiting magnitudes are 17.1, 16.4 and 15.3 mag in $J$ (1.25 nm), $H$ (1.65 nm), and $K_s$ (2.17 nm), respectively. Typical rms errors in proper motions range from better than 2 mas/yr for the brightest stars with Tycho-2 \citep{tycho2} data to 10 mas/yr or even 15 mas/yr in the region south of $\delta < -30^\circ$ \citep{ppmxl}. For the photometric data, the $1 \sigma$ uncertainty is typically better than 0.03 mag \citep{cat2MASS} for bright stars (< 13 mag) and increases up to 0.1 mag for fainter stars.

\subsection{Clusters and stars}

Although the main targets of our study are Galactic open clusters, we also considered such representatives of the Galactic population as stellar associations, embedded clusters, and globular clusters, which can also be identified in 2MAst.

\subsubsection{Compiling the cluster list}

The target list of 3784 entries with initial cluster parameters was compiled from sources available in the literature. As the primary source we used the data from the COCD \citep{clucat,newc109}, which was given the highest priority. For more optical clusters the data were taken from the \citet{daml02} list (Version 3.1, 24/nov/2010). Known associations were retrieved from \citet{melcat09}. For clusters detected in the NIR the information came from \citet{bicaea03}, \citet{dubisb03}, \citet{froeb07}, \citet{froea10}, and \citet{bukea11}. Globular clusters were selected from the catalogue by \citet{harris96}[edition 2010]\footnote{http://www.physics.mcmaster.ca/resources/globular.html}. Additionally, we incorporated data on radial velocities from \citet{newrv07} as well as supplementary data on embedded clusters from \citet{bicadb03} and \citet{lala03} and on stars in associations from \citet{humphr78}. The statistics of the available data is given in Table~\ref{tbl:cl_list}. We note that the cluster parameters in Table~\ref{tbl:cl_list} are highly heterogeneous because they are based on different kinds of observations as well as on different methods of membership defintion and parameter determination.

The compiled list of targets was sorted with increasing right ascension and numbered from MWSC~1 to MWSC~3784. We kept this identification even if the coordinates of cluster centres turned out to be changed in the process of the study.

\begin{table}
\caption{Statistics on available data for star clusters in the Milky Way}
\label{tbl:cl_list}
\begin{center}
\begin{tabular}{lc}
\hline
\noalign{\smallskip}
Parameter&Number \\
\hline
\noalign{\smallskip}
Number of entries    & 3784 \\
Angular size         &3648\\
Proper motion        & 957 \\
Radial velocity      & 730 \\
Distance             &2175 \\
Reddening            &1724 \\
Age                  &1562 \\
King tidal parameters&1830\\
\hline
\end{tabular} 
\end{center}
\end{table}

\subsubsection{Selection of areas and stars}

Around each cluster we initially defined a circular area with a radius of
$$
r_a = r_{cl} + r_{add}\,,
$$
where $r_{cl}$ is taken from the literature, and $r_{add} = 0\fdg3$. In a few cases, where $r_{cl}$ is unknown, we adopted $r_{cl}=r_{add}$ as a first approximation. If $r_{cl}$ and/or the coordinates of the cluster centre changed during the data processing, we repeated this step with new input values.

In the areas defined above we selected in 2MAst only those stars with flags $Rflg$ (the 2nd triple of the flags in 2MASS) set to 1, 2, or 3 in each band, i.e. the stars with the best-quality  detections in photometric and astrometric data. For a few stars spectral types are available from \ascc. These data were also incorporated since this additional information can be useful for the determination of cluster parameters.

\section{Construction of the pipeline for cluster parameter determination}
\label{sec:pipeline}

The main task of the pipeline is separating of cluster stars from the field by use of kinematic, photometric, and spatial criteria and determining cluster parameters from the data of the most probable cluster members. The pipeline is conceived to work iteratively, and, in principle, its philosophy is very similar to what we used in \citet[]{clucat}. However, a switch from the optical ($B$, $V$) photometry to the NIR ($J$, $H$, $K_s$) photometry required considerable modifications of the pipeline.

\subsection{Isochrones and reference sequences in colour-magnitude diagrams (CMD)}
\label{sec:iso}

The NIR isochrones for the 2MASS photometric system were computed using the Padova web-server CMD2.2\footnote{http://stev.oapd.inaf.it/cgi-bin/cmd}, based on the \citet{marigo08} calculations for $Z=0.019$. To improve the agreement of the lower part of the ZAMS with observations, we substituted the isochrones in the domain of late-K$-$M-stars by recent calculations of the Pisa group \citep{ndario10,moroni10}, taking also into account the latest Allard models\footnote{http://perso.ens-lyon.fr/france.allard/} of M-dwarf atmospheres. The pre-main sequence (Pre-MS) portion of the isochrones for $\log t\le 8.0$ was interpolated from Pre-MS tracks provided by Lionel Siess' web-server\footnote{http://astropc0.ulb.ac.be/~siess/database.html} \citep{siessea00} for $Z=0.02$. The derived $\log g$, $\log L/L_\odot$  and $T_{eff}$ were then transformed to the $JHK_s$-system with help of the Padova transformation tables.

We also used theoretical isochrones to define the zero-age main sequence (ZAMS) and terminal-age main sequence (TAMS) sequences in the CMD. For the brighter part (earlier than K-type), the ZAMS is represented by the Padova isochrone for $t=0.001$ Myr, whereas for the fainter part, the  Pisa Pre-MS isochrone for $t=650$ Myr and $(Y,Z)=(0.28,0.016)$  was chosen. The latter provides a better agreement with observations of Hyades M-dwarfs \citep{hyades11}. Thus, at low luminosities, the ZAMS defines a lower envelope of the cluster MS and coincides with
the locus of evolved Pre-MS stars at the age of 650 Myr. To avoid a discontinuity due to the use of different models, we corrected the Pisa isochrone, assuming that both sequences must demonstrate the same
mass-luminosity relation in the $\log m/m_\odot=-0.1\dots0.05$ mass range. The respective corrections are quite small,  $\Delta M_{K_s}=0.11$, $\Delta (J-K_s)=0.018$. As a result we have a smooth ZAMS. The TAMS sequence is constructed from models of stellar evolution dating the end of the hydrogen-burning phase in the cores. Both curves are shown in the $K_s$, $(J-K_s)$ diagram in Fig.~\ref{fig:atlp2} in Appendix C.

\subsection{Interstellar extinction}
\label{sec:extinction}

The three-band photometry from 2MASS allows one to determine the parameters of interstellar extinction from the photometric data only. The corresponding relations were taken from the average extinction law of \citet{card89}, whereas the transformation of $K$-magnitudes to $K_s$ was taken from \citet{dubi01}:
\begin{eqnarray}
A_K/A_{K_s} &=& 0.95\,,\nonumber\\
A_V/E(B-V)&=& 3.1\,,\nonumber\\
E(J-K_s)/E(B-V) &=& 0.480\,,\\
A_{K_s}/E(J-K_s) &=& 0.670\,.\nonumber
\end{eqnarray}
The reddening-free parameter $Q_{JHK_s}$ was adopted in the form
\begin{equation}
Q_{JHK_s}=(J-H)- \chi\,(H-K_s)\,,
\end{equation}
where the slope of the reddenning line $\chi$ is in first approximation equal to the ratio of the colour excesses
\begin{equation}
\chi \approx E(J-H)/E(H-K_s)\,.
\end{equation}
Apart from parameters of the photometric system, $\chi$ depends on the stellar spectral type, on properties of the absorbing matter along the line of sight (i.e. on the extinction law), and on the extinction value itself. Therefore, it is desirable to derive the slope directly from observations whenever possible. Based on earlier publications, \citet{mathis90} showed that the ratio $E(J-H)/E(H-K)$ varies between 1.61 and 2.09, but prefered the former value. \citet{strala08} have found from observations of red-clump giants in the inner Galaxy that for extinction $A_V<12$ the slope is close to $\chi$$=$2. Before accepting this value, we tested the effect of variations of $\chi$ and found that only $\chi=2$ provides independence of $Q_{JHK_s}$ from the reddening. Lower values lead to an increase of $Q_{JHK_s}$ with $E(J-K_s)$, while higher values decrease $Q_{JHK_s}$.

Before applying the above relations we tested them with help of available photoelectic UBV-photometry in open clusters located in different directions of the Milky Way and found no bias between our determinations and observations. The subsequent comparison with literature data carried out in Sect.\ref{sec:phopar} has confirmed the correctness of this procedure.

\subsection{Membership probabilities}
\label{sec:memb}

Following the approach applied earlier for the COCD clusters \citep{starcat}, we computed both kinematic and photometric membership probabilities for stars within the cluster areas. They were determined by taking into account the accuracy of the proper motions and of the stellar magnitudes.

We define the kinematic probability $P_{kin}^i$ of the $i$th star to belong to a cluster as
\begin{equation}
P_{kin}^i =  \exp \left\{-\frac{1}{4}\left[
                 \left(\frac{\mu_x^i - \overline{\mu}_x}
                               {\varepsilon_{\mu^i}}\right)^2 +
                 \left(\frac{\mu_y^i - \overline{\mu}_y}
                               {\varepsilon_{\mu^i}}\right)^2
                 \right]\right\}, \label{ppm_eqn}
\end{equation}
where $\mu^i_{ x,y}$ and $\overline{\mu}_{x,y}$ are the components of the proper motion of the $i$th star and the mean proper motion of the cluster, respectively. The parameter $\varepsilon_{\mu^i}$ was set to 1.5~mas/yr for each star that had mean errors of proper motion smaller than this threshold in 2MAst. For all other stars, $\varepsilon_{\mu^i}$  corresponds to the mean proper motion errors given in 2MAst.

The photometric probabilities $P_{JH}$ and $P_{JK}$ are computed in two CMDs, $K_s,\,(J-H)$ and $K_s,\,(J-K_s)$, respectively. Again, we followed the approach developed in \citet{starcat} for the photometric selection of cluster members. For each isochrone we defined the most probable location of cluster members as an area in between the corresponding smoothed isochrones for single stars and for unresolved binaries of equal mass. It follows that at a given magnitude $K_s$ there exist two values of colour $c_b$ (blue) and $c_r$ (red), which limit this domain in a CMD. A star $i$ with a colour $c^i$ is assumed to be a photometric member with a probability $P^i_c=100$\% if $c_b\le c^i\le c_r $. For the remaining stars we computed the photometric probability as
\begin{equation}
P^i_c = \exp \left\{ -\frac{1}{2}\left[
               \frac{\Delta c^i}{\varepsilon_{c}^i}\right]^2
\right\}, \label{pph_eqn}
\end{equation} 
where  $\Delta c^i = c^i -c_b$ for $c^i < c_b$, and $\Delta c^i=c^i - c_r$ for $c^i > c_r$. The quantity  $\varepsilon_{c}^i$ is the mean error of the colour index $c^i$. It is calculated from the individual $rms$ errors of the magnitudes. A minimum $rms$ error of a magnitude of 0.03~mag is assumed. In this way we obtained two photometric probabilities, $P_{JH}$ and $P_{JK}$ for each star.

Finally, we defined a combined probability $P$ that takes into account all aspects of the membership selection procedure, i.e. kinematic, photometric, and spatial selection criteria ($P_s$)
\begin{equation}
P =P_s\cdot\mathrm{min}\{P_{kin},P_{JH},P_{JK}\}\,,
\end{equation}
where the factor  $P_s$ is set to 1 within the cluster radius $r_2$, and to 0 otherwise (see Sect. 3.4.1). The probabilties $P_{kin},P_{JH},P_{JK},P_s$ are given in the catalogue of cluster stars MWSCS, and the user may feel free to define his/her own selection criteria.

As in \citet{starcat}, we classified the quality of cluster membership. We considered three classes of cluster members ($1\sigma$-members as the most probable cluster stars, $2\sigma$-members as possible cluster stars, $3\sigma$-members as possible field stars). Logically this classification is expressed by the following constraint: $P_s=1\,\wedge\,P_{kin} > p_0\,\wedge\, P_{JK} > p_0 \,\wedge\, P_{JH} > p_0$, where the threshold probability $p_0=$ 61\% for $1\sigma$-, 14\% for $2\sigma$- and 1\% for $3\sigma$-members.

\subsection{Cluster parameter determination}\label{sec:pardet} 

In general, the parameter determination follows the rules applied earlier to the clusters in the COCD \citep{clucat}. However, due to the considerably fainter limiting magnitude, 2MAst provides many more cluster members. In contrast to COCD-clusters we can, therefore, determine the cluster parameters from the most probable members ($1\sigma$-members) located in the central cluster areas ($r<r_1$, see  Sect.\ref{sec:spapar} for the definition of $r_1$), where the contamination by field stars is minimal. Consequently, the reliability of the results could be improved. Furthermore, the stars included in the solution had to fulfill the following conditions:
\begin{itemize}
\item the 2MASS flag Qflg is "A" (i.e., signal-to-noise ratio $S/N>10$) in each photometric band for stars fainter than $K_s=7.0$;
\item the mean errors of proper motions are smaller than 10 mas/yr for stars with $\delta \geq -30^\circ$, and smaller than 15 mas/yr  for $\delta < -30^\circ$.
\end{itemize}

All parameters were determined in a process 
aiming at compiling a consistent list of cluster members 
that in turn was used for revising the cluster parameters. 
This led to  a series of iterations that each consisted of the following steps:
\begin{enumerate}
 \item based on the parameters derived in the previous iteration, we computed for all stars in the cluster area the four membership probabilities discussed above ($P_s,\,P_{kin},\, P_{JH},\, P_{JK}$);
\item based on the computed values of the probabilities, we constructed a corrected list of cluster members and produced the spatial, kinematic, and photometric diagrams (shown in Figs.~\ref{fig:atlp1},\ref{fig:atlp2});
\item visual inspection and analysis of the diagrams for the determination of new cluster parameters: cluster coordinates and radii, average proper motion, age, distance, and reddening (as described in Sect.~\ref{sec:spapar}, \ref{sec:kinpar}, and \ref{sec:phopar}).
If these parameters did not agree with those from the previous iteration and/or the member list differed from the previos one, the parameter list was corrected and the iteration was repeated, otherwise the process was stopped.
\end{enumerate}
As initial cluster parameters we used data taken from the literature. 
If no literature data were available for a cluster, we set cluster parameters to values derived from an initial visual inspection of the data
in the cluster field and to typical cluster parameters (ages, distances).

To keep the derived parameters homogeneous and following the COCD practice, the pipeline was run by and under control of a single team member (NVK). A cluster was considered as ``confirmed'' by the pipeline when compatible and non-contradictory data on cluster structure, kinematics, and CMD were derived, and when the number of the most probable cluster members was found to be not less than five. Objects that did not fulfil these criteria were considered as ``not confirmed''.

\begin{figure}[t]
\includegraphics[width=\hsize,clip=]{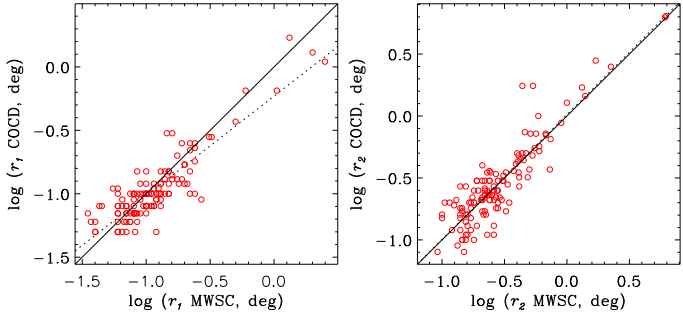}
\caption{Comparison of apparent radii of open clusters (left panel for $r_1$,
right panel for $r_2$) derived with corresponding data from the COCD. The straight line is a bisector, the dotted line is the linear fit of the data.}
\label{fig:comr12}
\end{figure}

\subsubsection{Cluster identification and spatial parameters of clusters}\label{sec:spapar}

Spatial cluster parameters were determined from stellar density profiles obtained from star counts. For each cluster, the counts were carried out in concentric circles around the cluster centre with a minimum stepsize of $0\fdg008$. For visual inspection and check of cluster identification we used \citet{archy03} as well as the SIMBAD\footnote{http://simbad.u-strasbg.fr/simbad/} and WEBDA databases, CDS Portal\footnote{http://cdsportal.u-strasbg.fr/}, and DeepSkyBrowser\footnote{http://messier45.com/} resources. The density profiles were computed for four different stellar samples: $1\sigma$-, $2\sigma$-, and $3\sigma$-cluster members, and all stars. The centres were determined by eye as the points of maximum surface density of the most probable cluster members  (see Fig.~\ref{fig:atlp1} in Appendix~\ref{sec:atl} for illustration). Compared to the literature data, the coordinates of the cluster centres agree within 2.5~arcmin for more than 90\% of the clusters. However, for the remaining clusters the differences can reach $0\fdg1$ and more. This is usually the case for relatively nearby large-sized clusters or associations, observed with a rich or irregular stellar background. For some objects this occurs due to mistakes in the source catalogues.

The radial density profile (RDP) of the $1\sigma$-members was the decisive factor for the determination of the cluster size, though we also took into account their location in the cluster maps and the radial distributions of their $K_s$-magnitudes and proper motions (see Fig.~\ref{fig:atlp1} in Appendix~\ref{sec:atl} for illustration). We assumed a centrally  symmetric distribution of cluster members around the cluster centre and introduced three empirical structural parameters $r_0 <  r_1 < r_2$ describing the shape of the RDP, which were fitted by eye. The visible core radius $r_0$ corresponds to the distance from the cluster centre where the slope of the RDP becomes flatter, the visible radius of the central part $r_1$ is the distance where the decrease of the stellar density stops abruptly, and the actual (total) visible radius of a cluster $r_2$ is defined as the distance from the cluster centre where the surface density of stars becomes equal to the average density of the surrounding field. $r_0$ is a new feature considered here, while in our previous work with the COCD we determined only $r_1$ and $r_2$. Based on visual inspection of the corresponding data, $r_0$, $r_1$, and $r_2$ give empirical descriptions of cluster sizes, without a direct association to e.g. King's parameters $r_c$ and $r_t$ (see Sect. 3.4.4.). However, we found that the relations $r_0 < r_c < r_1$ and $r_t > r_2$ are generally obeyed.

One can judge the accuracy of the derived radii by comparing our determinations with COCD-data. In Fig.~\ref{fig:comr12} and Table~\ref{tbl:comp_rpm} we compare the radii $r_1$ and $r_2$ determined for MWSC and COCD clusters. One can see in Fig.~\ref{fig:comr12} that both values are well represented in logarithmic scale by a straight line
\begin{equation}
p(COCD) = a + b\,p(MWCI)
\end{equation}\label{eq:prmlf} 
\noindent 
with fitting parameters as shown in Table~\ref{tbl:comp_rpm}. Although the MWSC radii are based on a much deeper catalogue (2MAst), they agree reasonably well with the results obtained with \ascc data.

\begin{figure}[t]
\includegraphics[width=\hsize,clip=]{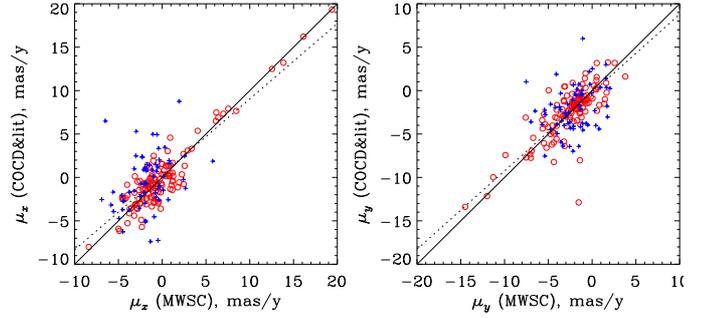}
\caption{Comparison of the average cluster proper motions (left panel for $X$-component, right panel for $Y$), derived in the present study with corresponding data from the literature. COCD clusters are shown by red circles, other clusters are shown by blue plusses. The straight line is a bisector, the dotted line is the linear fit of the data.
}
\label{fig:compm}
\end{figure}

\subsubsection{Kinematic parameters}\label{sec:kinpar}

For each cluster, the weighted mean components $\olin{\mu}_{x,y}$ of the cluster proper motion were computed from the proper motions of the most probable members ($1\sigma$-members) within $r_1$. Usually, we found 20-40 stars per cluster that fulfil this condition. Compared to COCD, the typical formal accuracy of cluster proper motions is improved (0.7 mas/yr vs 0.4 mas/yr). No systematic differences were found between literature and MWSC proper motions (see Fig.~\ref{fig:compm} and Table~\ref{tbl:comp_rpm}).

\begin{figure*}[t]
\sidecaption
\includegraphics[width=12cm]{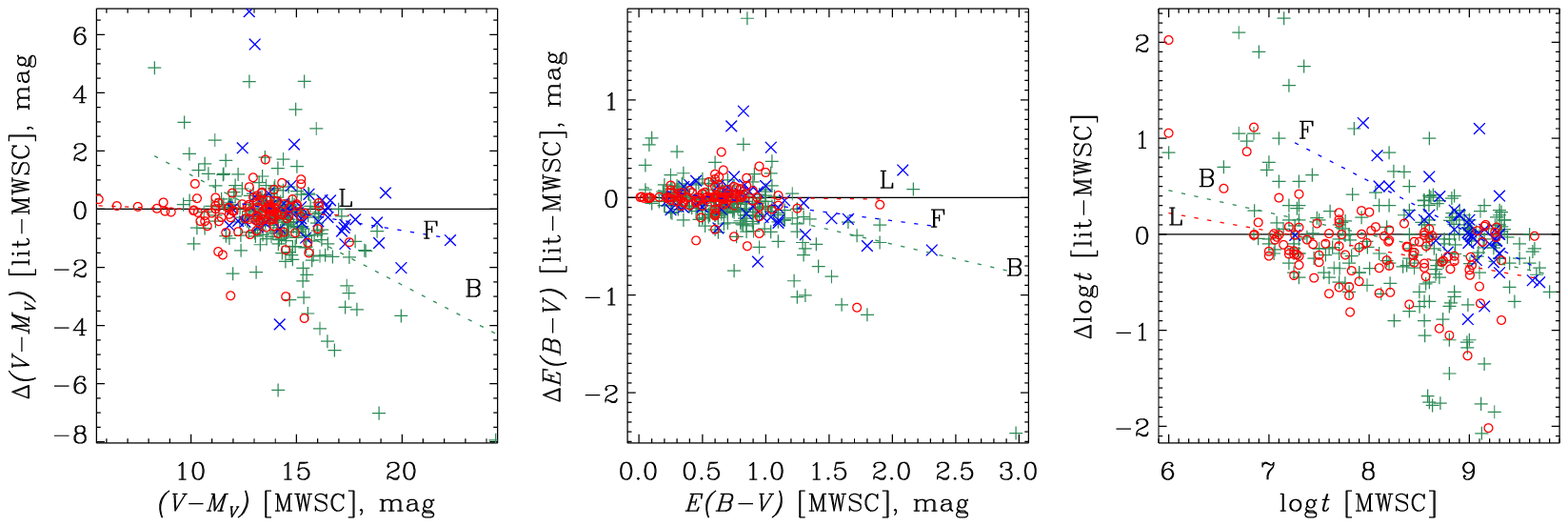}
\caption{Comparison of distance moduli (left panel), reddenings (middle panel), and ages (right panel) derived with literature data. Optical data \citep{loktin01,loktin04} are shown by red circles, the NIR data are marked by blue crosses \citep{froea10} and green plusses \citep{bukea11}. The dotted lines labelled L(Loktin), F(Froebrich), and B(Bukowiecki) are the corresponding fitted linear regressions (see Table~\ref{tbl:comp_ph} for their parameters).}
\label{fig:comph}
\end{figure*}

We also compared MWSC proper motions with data taken from the literature \citep[as collected in][]{daml02}. Basically these are the results of \citet[]{diaslt1,diasgt1} and \citet{beshenov}. The cluster proper motions agree reasonably well. A higher dispersion can probably be explained by the fact that the data in the papers cited above are heterogeneous compilations.

\begin{table}[b]
\caption{Fitting parameters for relations between MWSC and literature data for cluster radii (in deg) and proper motions (in mas/y).}
\label{tbl:comp_rpm}
\begin{tabular}{@{}lrrrr}
\hline
\noalign{\smallskip}
Parameter&$N$&\mc{1}{c}{a}&\mc{1}{c}{b}&St.dev.\\
\hline
\noalign{\smallskip}
$\log r_1$&126&$-0.23\pm0.04$&$0.78\pm0.04$&0.13 \\
$\log r_2$&126&$ 0.01\pm0.03$&$1.01\pm0.04$&0.15 \\
$\mu_x$   &189&$ 0.36\pm0.17$&$0.87\pm0.04$&1.54 \\
$\mu_y$   &189&$-0.23\pm0.19$&$0.90\pm0.03$&1.92 \\
\hline
\end{tabular}\vspace{1mm}\\
\end{table}

As a rule, average radial velocities of star clusters were adopted from \citet{daml02}, and \citet{harris96}. When possible we computed mean radial velocities of the clusters using data on single stars from the CRVAD2-catalogue \citep{newrv07}. Because there are few stars with known radial velocities, we had to include data on $1\sigma$, $2\sigma$ and sometimes even $3\sigma$-members distributed over the total area of a cluster within the radius $r_2$.

\subsubsection{Photometric parameters and ages of clusters}\label{sec:phopar} 

To determine the interstellar reddening, we used a colour-colour diagram $(H-K_s),\,(J-H)$, while the diagram $Q_{JHK_s},\,(J-K_s)$ was used for the control of the derived values. In the first diagram the reddening was estimated by a shift of the corresponding isochrone along the reddening vector from its intrinsic position until the best agreement with the observations was achieved. In the second diagram we applied a horizontal shift of the isochrone which occurs when $\chi$ is set to 2. Both diagrams are shown in Fig.~\ref{fig:atlp2} in Appendix~\ref{sec:atl} together with the initial and final positions of the isochrones. The distance to the cluster was determined by a shift of the isochrone in the colour-magnitude diagrams $K_s,\,(J-H)$ and $K_s,\,(J-K_s)$. However, the choice of a proper isochrone presumes knowledge of the cluster age. Therefore, the cluster distance, age, and reddening were determined in an iterative procedure. The iterations were stopped when the parameters did not change anymore.

\begin{table}[b]
\caption{Comparison of MWSC and literature data for photometric parameters and age. See text for the explanation
of the parameters}
\label{tbl:comp_ph}
\begin{tabular}{@{}lrrrr@{\extracolsep{2mm}}c@{}}
\hline
\noalign{\smallskip}
Parameter&$N$&\mc{1}{c}{a}&\mc{1}{c}{b}&St.dev.&Source\\
\hline
\noalign{\smallskip}
$(V-M_V)$&109&$-0.06\pm0.05$&$-0.03\pm0.05$&0.48 &L\\
$(V-M_V)$&43 &$ 0.23\pm0.13$&$-0.12\pm0.13$&0.64 &F\\
$(V-M_V)$&227&$ 0.42\pm0.08$&$-0.38\pm0.08$&0.93 &B\\
$E(B-V)$ &109&$ 0.00\pm0.01$&$-0.01\pm0.01$&0.08 &L\\
$E(B-V)$ &43 &$ 0.02\pm0.05$&$-0.17\pm0.05$&0.21 &F\\
$E(B-V)$ &227&$-0.03\pm0.02$&$-0.30\pm0.02$&0.16 &B\\
$\log t$ &109&$-0.15\pm0.03$&$-0.18\pm0.03$&0.31 &L\\
$\log t$ &43 &$ 0.55\pm0.09$&$-0.54\pm0.09$&0.23 &F\\
$\log t$ &227&$-0.00\pm0.04$&$-0.23\pm0.04$&0.49 &B\\
\hline
\end{tabular}\vspace{1mm}\\
References: L=\citet{loktin04};F=\citet{froea10};B=\citet{bukea11}
\end{table}

We found that the theoretical isochrones usually fitted the observed sequences in the $K_s,(J-K_s)$-diagrams better than in $K_s,(J-H)$. We observed a $(J-H)$ colour shift that is typically of about a few hundreds of magnitude but it varied from cluster to cluster. We attribute this effect to a lack of a global calibration of the $H$-band in 2MASS \citep[see][]{cat2MASS} and, additionally, to spatial variations of properties of absorbing dust that interfere in the $H$-band. Therefore, we introduce an empirical correction $\Delta H$ to provide a better isochrone fit in the $K_s,(J-H)$ diagram.

To determine the ages of older clusters we applied the procedure described in \citet{clucat} where  the average age of the turn-off stars is used as a cluster age indicator. This approach is justified for evolved clusters with well-defined turn-off points in the CMD (it applies for 33\% of the considered clusters). In a younger cluster one observes a relatively poorly populated upper MS, which is augmented sometimes by a Pre-MS branch if the cluster is sufficiently close to the Sun.  Therefore, the age determination of younger clusters was carried out with the isochrone-fitting technique. In both methods we used the $K_s, (J-K_s)$-diagrams built from the most probable members ($1\sigma$-members) located in the central cluster areas ($r<r_1$). The selection of the best-fitting isochrone was made by eye.

Comparing the newly derived distances and ages with COCD data, we find that both systems coincide for distances ($\langle\Delta d\rangle=4\pm25$ pc) and show a small bias for ages ($\langle\Delta \log t\rangle = 0.08\pm0.03$). A comparison of the derived parameters with published data is shown in Fig.~\ref{fig:comph} and Table~\ref{tbl:comp_ph}, where we discuss the results of \citet{loktin01}, \citet{loktin04}, \citet{froea10}, and \citet{bukea11}. The \citet{loktin01} and \citet{loktin04} results are based on  photoelectric UBV observations collected in WEBDA and provide the most extended and homogeneous set of cluster parameters obtained from optical photometry. Since our measurements and those of \citet{loktin01} can be regarded as independent, we can consider the respective differences as a measure of accuracy of the derived values. Assuming that the data have approximately equal weight, we find that our distances are accurate to within 16\% (with a typical error in the distance modulus of $\pm$0.35~mag) and our reddening values $E(B-V)$ are accurate to within $\pm$0.06~mag (or about 10\%). We note that for young clusters, which have almost vertical and poorly populated main sequences, the accuracy of the parameters is considerably lower. The results from \citet{froea10} and \citet{bukea11} were obtained with 2MASS data. Note that all these studies are based on a photometric selection of cluster members.

To check possible systematic differences between our results and the published parameters, we constructed a linear regression
\begin{equation}
\Delta p = p(\mathrm{lit}) - p(\mathrm{MWSC}) = a + b\,\,[p(\mathrm{MWSC})-p_0]\,,
\end{equation}
where $p$ is one of the parameters ($E(B-V)$, $(V-M_V)$, or age $\log t$), and $p_0$ is a constant taken as 0.5~mag, 12~mag, and 8 for these parameters respectively. Concerning the distance moduli and reddenings, the results generally agree reasonably well with optical \citep{loktin01, loktin04} and with NIR data \citep{froea10}, but they show a systematic difference to the data of \citet{bukea11}, though the scatter of the data points is rather large.

The comparison of ages points to a more serious problem that probably arises from the different approaches of membership determination. Indeed, a purely photometric selection usually favours brighter stars as cluster members. However, they can be contaminating field stars, and their membership can be rejected by kinematic constraints. With such stars included as members, the cluster ages are underestimated, and the difference to MWSC ages becomes negative, as seen in Fig.~\ref{fig:comph} for higher MWSC ages. On the other hand, when brighter members are excluded in the literature, the resulting age  becomes too old compared to our estimate. Therefore, the membership question has a direct impact on the age determination of clusters, and this can be responsible for the  discrepancies in the determination of $(V-M_V)$ and $E(B-V)$, too. Because our membership is based on kinematic and photometric constraints, we believe that our results are more reliable. Another reason for overestimated ages of young clusters can be the neglect of Pre-MS stars in the age determination (a procedure typical to some recent open cluster surveys), although they fix younger ages more reliably than MS-stars can do. As our analysis shows, all noticeable positive deviations of the \citet{bukea11} clusters in Fig.~\ref{fig:comph} can be explained by the reasons above.

\begin{figure}[t]
\includegraphics[width=0.975\hsize,clip=]{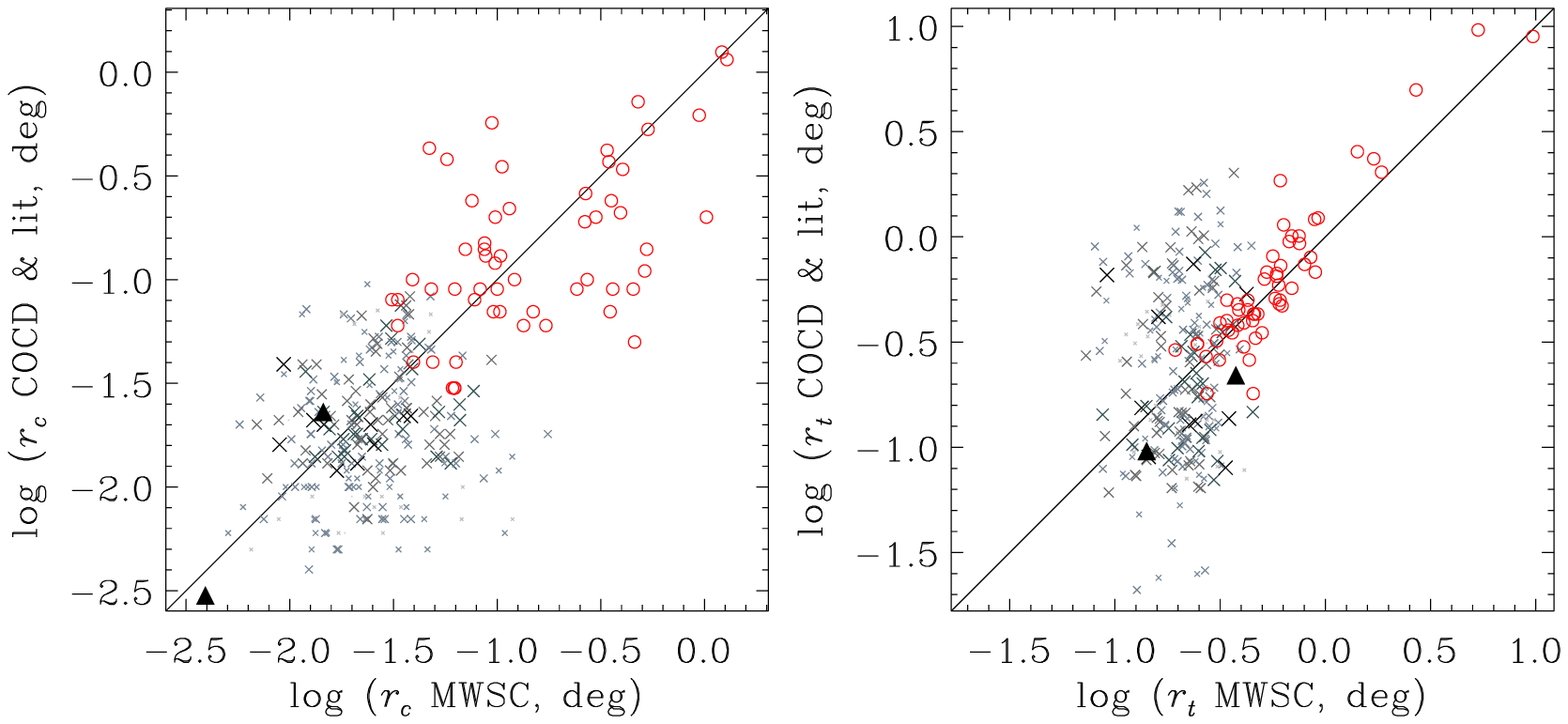}
\caption{Comparison of tidal parameters of open clusters (left panel for $r_c$, right panel for $r_t$), derived with literature data. COCD clusters \citep{clumart} are marked as red circles. \citet{froeb07} data are shown by crosses. Their size and intensity is proportional to the quality flag in \citet{froeb07}. Larger sizes and darker colours correspond to better quality. Two filled triangles show globular clusters identified in the 2nd quadrant. Their tidal parameters are taken for comparison from the 2010 edition of the catalogue of \citet{harris96}.}
\label{fig:comprt}
\end{figure}

\subsubsection{Tidal parameters}

After completing the iterative parameter determination described in the previous
subsections, we determined the tidal parameters $r_c$  (core radius), $r_t$ (tidal radius) using the same technique as applied earlier to COCD clusters \citep[see][for details]{clumart}. The method is based on a three-parameter fit of cumulative King profiles to the observed density distribution of cluster members. Since 2MAst is much deeper than the \ascc catalogue, a considerably better member statistics in a MWSC cluster is now available. Already from this point of view we expect a better random accuracy of the tidal parameters. However, it is not a priori clear whether the new and old determinations do not differ systematically.

In Fig.~\ref{fig:comprt} we compare the newly derived radii $r_c$ and $r_t$ with our previous determinations (red open circles in Fig.~\ref{fig:comprt}). Though the data show a relatively large scatter for core radii $r_c$, we found no systematic differences between the new and old determinatations. Also, the tidal radii $r_t$ derived with \ascc and 2MAst agree well for the majority of clusters. For a few nearby clusters ($r_t >$ 1~deg), the COCD tidal radii seem to be slightly overestimated. However, we prefer to postpone the discussion on the significance of a possible bias until at least the next quadrant is processed.

\begin{figure*}[t]
\includegraphics[width=\hsize]{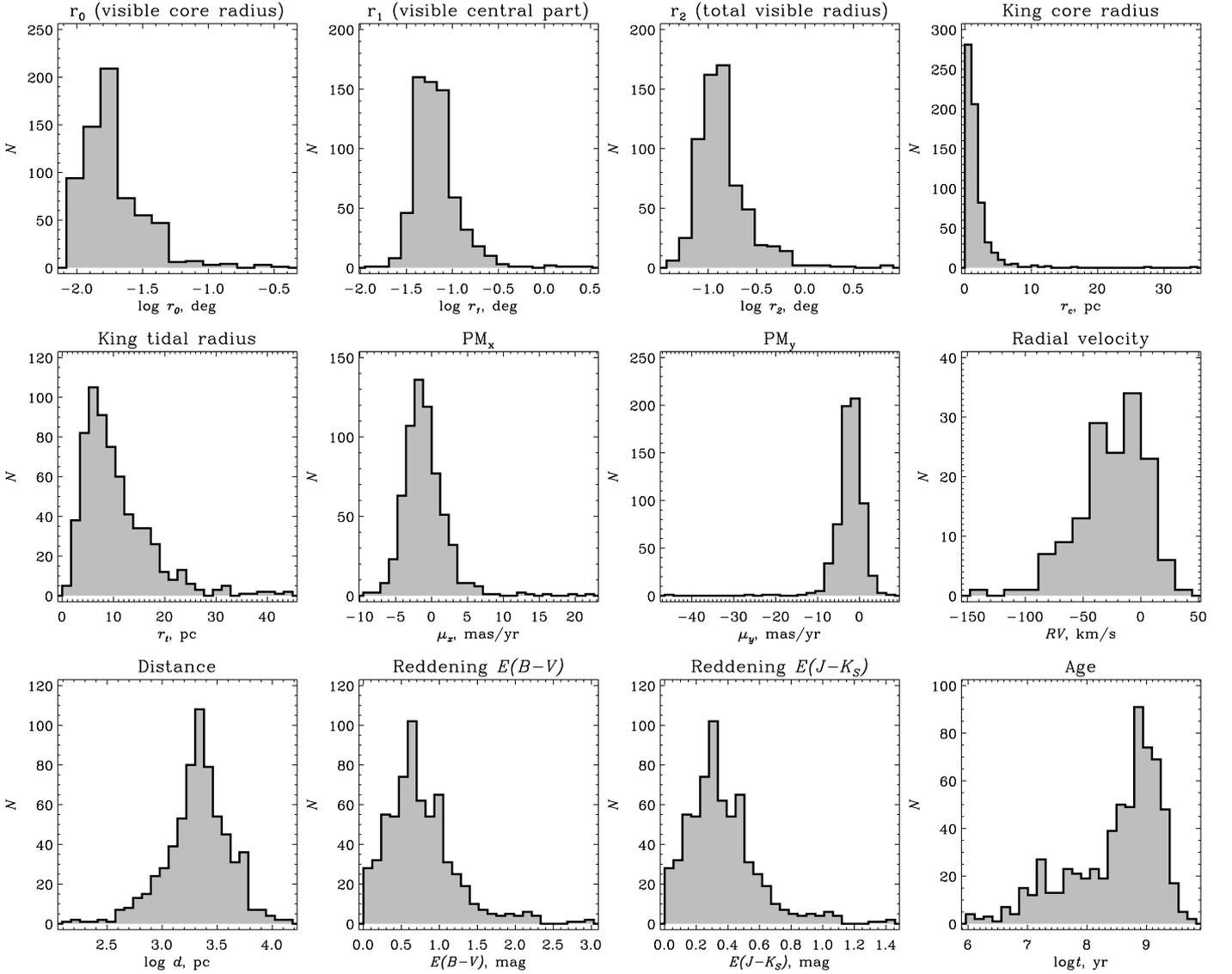}
\caption{Distribution of star clusters in the second quadrant over derived fundamental parameters.
}
\label{fig:dstr_all}
\end{figure*}

In Fig.~\ref{fig:comprt} we also show the tidal parameters determined by \citet{froeb07}, who obtained cluster radii by fitting King profiles to the observed overdensities of 2MASS stars, neglecting any kinematic and/or photometric selection of members. The comparison with \citet{froeb07} is shown with crosses in Fig.~\ref{fig:comprt}, where the size and intensity of the crosses correspond to the quality level of the determinations as given by the authors. Larger crosses and darker colour describe a higher quality of the results. One can see that the clusters of \citet{froeb07} exhibit systematically smaller angular core and tidal radii than the COCD clusters. This systematic displacement simply reflects the fact that the clusters detected by \citet{froeb07} in the near-infrared are on average farther away from the Sun than the optical COCD clusters, which represent the local cluster population. Considering only the \citet{froeb07} sample, we do not see a correlation of the results in both panels of Fig.~\ref{fig:comprt}, probably due to the large relative errors. However, the higher dispersion of the tidal radii of \citet{froeb07} is in our opinion caused by a lower accuracy of their determination. This is not surprising since \citet{froeb07} fitted King curves to observed profiles contaminated by field stars. This introduced additional noise in their determination. Note that the clusters with the largest discrepancies in both core and tidal radii (Fig.~\ref{fig:comprt}) were marked as of lower quality by \citet{froeb07}, who also mentioned a generally much lower accuracy in their determination of tidal radii compared to the core radii.

In Fig.~\ref{fig:comprt} we added comparisons for a few globular clusters shown as filled triangles. Out of four globular clusters located in the second quadrant we were able to fit density profiles for two objects (NGC~288 and Pal~2). The other two clusters (Pal~1 and Whiting~1) are too faint to show reasonable profiles in 2MAst. We compared our results with those from the catalogue of the Galactic Globular cluster parameters \citep[][2010 edition]{harris96}. The tidal parameters from the literature coincide sufficiently well with our determinations.

To give the reader an overview of the properties of the open star clusters in the second quadrant, we show in Fig.~\ref{fig:dstr_all} the distributions of the cluster parameters.

\section{Star cluster population in the 2nd quadrant of the Milky Way}
\label{sec:clupop}

The current survey includes 652 objects (642 open, 2 globular clusters, and 8 associations) or 76\% out of 871 from the input list, that could be confirmed as clusters in the sector of Galactic longitudes $l=90^\circ\dots180^\circ$. About half (122) of the rejected targets turned out to be random density enhancements of field stars, the others (97) are too faint to be detected in 2MAst.

\begin{figure}[t]
\includegraphics[width=\hsize]{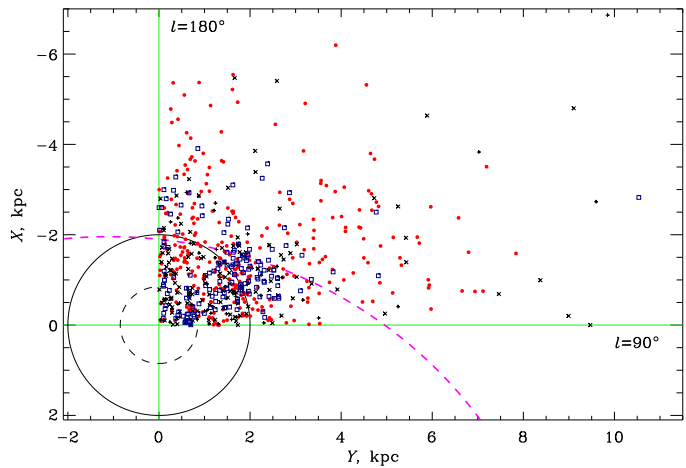}
\caption{Distribution of star clusters in the $XY$-plane centred at the position of the Sun. Blue squares mark the youngest ($\log t \leq 7.9$) clusters,  black plusses show clusters with $7.9 < \log t \leq 8.3$, black crosses correspond to clusters with $8.3 < \log t \leq 8.75$, while red circles represent old clusters with $\log t > 8.75$. The dashed magenta thick curve is the Perseus spiral arm fitted by a two-arm model of the spiral pattern with a pitch-angle of $-6^\circ$. The large black dashed circle around the Sun indicates the completeness area of the COCD sample. The  black solid circle marks the estimated probable completeness area of MWSC.
}
\label{fig:xy90_135}
\end{figure}

\begin{figure}[t]
\includegraphics[width=\hsize,clip=]{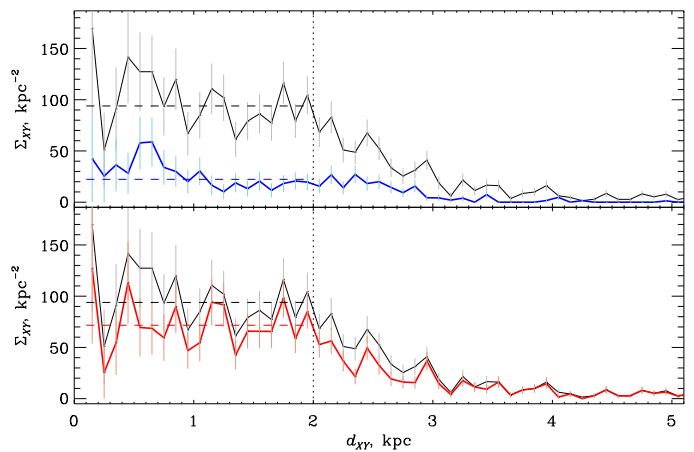}
\caption{Distribution of the surface density $\Sigma$ of star clusters in the 2nd Galactic quadrant versus their distance $d_{XY}$ from the Sun projected onto the Galactic plane. The density distribution of all clusters is given in black in both panels. In the upper panel we show the distribution of young clusters ($\log t<7.9$) in blue, in the lower panel that of older clusters($\log t\geq7.9$) in red. The dotted lines indicate the completeness limit, the dashed horizontal lines correspond to the average density at $d_{XY} \leq$ 2~kpc for a  given group of clusters.}
\label{fig:dcompl}
\end{figure}

For each confirmed cluster we provide the basic set of cluster parameters (see Fig.~\ref{fig:dstr_all} and Appendix). In Fig.~\ref{fig:xy90_135} we show the distribution of the 652 clusters projected onto the Galactic plane $(X,Y)$. For comparison, we also indicate the position of the grand design Perseus spiral arm with a pitch-angle of $-6^\circ$ fitted to the young COCD clusters by \citet{clupop}. To estimate the completeness of the MWSC cluster sample, we computed the surface density $\Sigma$ of the clusters as a function of the projected distance from the Sun. The result is shown in  Fig.~\ref{fig:dcompl} for 650 open clusters and associations as well as for two subsamples of clusters, i.e. younger and  older than $\log t = 7.9$. Although the distribution does reveal fluctuations, it can be considered as flat up to $d_{XY}\approx$ 2.0~kpc. Then, the distribution drops, and the cluster density decreases slowly with lager distances. This can be interpreted as an indication of the completeness limit of our survey at $d_{XY}\approx$ 2.0~kpc. The completeness limit is defined by older clusters, while we observe an almost uniform distribution of younger clusters in the range 2~kpc $< d_{XY}<$ 3~kpc, and a lower surface density only at larger distances. Probably, 3~kpc is the completeness limit for young clusters in our survey. However, we cannot exclude that the decrease in the density beyond about 3~kpc is a physical effect and related to the Perseus arm. A more conclusive answer is expected once the survey will be extended to $l > 180^\circ$.

Another interesting feature in the radial distribution of young clusters is a density peak at $d_{XY}\approx 0.6$  kpc. This density increase can be seen also in Fig.~\ref{fig:xy90_135} at $l\approx 90^\circ$ and is rather a footprint of the star forming region in Cygnus than a mere density fluctuation.

Considering the clusters within the completeness limit ($d_{XY}\approx$ 2~kpc), we found that the average surface density $\Sigma$ is 94 kpc$^{-2}$ in the 2nd quadrant, with  contributions of 22 kpc$^{-2}$ and 72 kpc$^{-2}$ from clusters younger and older than  $\log t = 7.9$, respectively. This estimate is somewhat lower than the $\Sigma = 114\, \mathrm{kpc}^{-2}$, which we obtained with COCD clusters in a smaller area within $d_{XY} < 0.85$ kpc \citep[see][]{clupop}.

\section{Summary}
\label{sec:conc}

We described the technique we applied to provide a global survey of open clusters in the Milky Way and reported the first results achieved in the 2nd Galactic quadrant. The target list was compiled from present-day lists of known open clusters and cluster candidates but it also includes associations, moving groups, and globular clusters. As the observational basis, we constructed a dedicated catalogue, called 2MAst, of about 470~million stars with proper motions from PPMXL and photometry from 2MASS. We developed a processing pipeline that provides simultaneous determination of the kinematic and photometric membership as well as estimating cluster parameters. The basic set of cluster parameters includes structure parameters (coordinates of the cluster centre, sizes), kinematic parameters (average proper motions), photometric parameters (colour excesses, extinction in the $K_s$ band, distance, and age).

In total, 871 objects were examined in the 2nd quadrant (with 850 cluster-like objects and 21 stellar associations). We confirmed 652 of them (76\%). Among them there are open and globular clusters, cluster-like (compact) associations, moving groups  and remnant clusters. We had to reject 12 large associations with sizes larger than a few degrees and with more than one peak in the density distribution of members over the projected sky area. For these ``multi-centre'' associations our pipeline failed to provide proper cluster parameters. Moreover, there were simply duplicate entries, or clusters too faint to be properly studied in 2MAst. The majority of the rejected clusters, however, were random clusterings of field stars.

The relatively high percentage of rejected clusters can be explained both by the ``risky'' character of the input list, which includes all possible types of objects (including those that are treated as asterisms in the literature), and as a consequence of the enhanced control of the results, where the automated pipeline is extended by human decision at every stage.

We determined the basic set of cluster parameters for all 652 confirmed objects and were able to determine tidal parameters via a direct fit of King profiles to observed density distributions of cluster members. Radial velocities could be provided for 151 objects (or 23\% of the confirmed clusters). For the first time, distance and age estimates are given for  291 and 333 clusters, respectively.

\begin{acknowledgements}
This study was supported by DFG grant RO 528/10-1, and RFBR grant 10-02-91338, and by Sonderforschungsbereich SFB 881 "The Milky Way System" (subproject B5) of the German Research Foundation (DFG). We thank N. Da~Rio and P.G.P. Moroni for providing their isochrones for Hyades-age clusters prior to publication. We acknowledge the use of the Simbad database, the VizieR Catalogue Service and other services operated at the CDS, France, and the WEBDA facility, operated at the Institute for Astronomy of the University of Vienna. We thank the anonymous referee for her/his extensive comments that enabled us to make the paper more easily readable.
\end{acknowledgements}

\bibliographystyle{aa}
\bibliography{clubib}

\appendix

\section{MWSCS.Q2: The Catalogue of Milky Way Cluster Stars in the 2nd Galactic
Quadrant}\label{sec:appscat}

\begin{table}[ht]
\caption{Contents of the MWSCS table of stellar data}
\label{tbl:stardat}
\setlength{\tabcolsep}{2pt}
\begin{tabular}{rlcl}
\hline
Col& Label&Units&Explanations\\
\hline
\noalign{\smallskip}
 1  &RA         &  h    &  Right ascension J2000.0, epoch 2000.0\\
 2  &DE         & deg   &  Declination J2000.0, epoch 2000.0\\
 3  &$B$        & mag   &  $B$ magnitude in Johnson system\\
 4  &$V$        & mag   &  $V$ magnitude in Johnson system\\
 5  &$J$        & mag   &  $J$ magnitude 2MASS\\
 6  &$H$        & mag   &  $H$ magnitude 2MASS\\
 7  &$K_s$      & mag   &  $K_s$ magnitude 2MASS\\
 8  &$\varepsilon_J$      & mag   &  Error of $J$ magnitude\\
 9  &$\varepsilon_H$      & mag   &  Error of $H$ magnitude\\
10  &$\varepsilon_{K_s}$  & mag   &  Error of $K_s$ magnitude\\
11  &PMx        & mas/yr &  Proper motion in RA*cos(DE)\\
12  &PMy        & mas/yr &  Proper motion in DE\\
13  &$\varepsilon_{PM}$   & mas/yr &  Error of proper motion\\
14  &$RV$       & km/s   &  Radial velocity\\
15  &$\varepsilon_{RV}$   & km/s   &  Error of RV\\
16  &Qflg       & ---    & 2MASS (ph\_qual) $JHK_s$ photometric\\
    &           &        & quality flag\\
17  &Rflg       & ---    & 2MASS (rd\_flg) source of $JHK_s$\\
    &           &        & default mag \\
18  &Bflg       & ---    & 2MASS (bl\_flg) $JHK_s$ components\\
    &           &        & fit to source\\
19  &2MASS      & ---    & 2MASS (pts\_key) unique source\\
    &           &        & identifier in catalogue 2MASS\\
20  &ASCC       & ---   & \ascc number\\
21  &Sp         & ---   & Spectral type and luminosity class\\
22  & $R_{Cl}$  & deg   & Angular distance from the adopted\\
    &           &       & cluster centre\\
23  & $P_{s}$   & ---   & Spatial probability of cluster\\
    &           &       & membership\\
24  & $P_{kin}$ &  \%   & Proper motion probability of cluster\\
    &           &       & membership\\
25  & $P_{JK_s}$&  \%   & Photometric $JK_s$ probability of cluster\\
    &           &       & membership\\
26  & $P_{JH}$  &  \%   & Photometric $JH$ probability of cluster\\
    &           &       & membership\\
27  &No         &  ---  & MWSC number\\
\noalign{\smallskip}
\hline
\end{tabular}\\
\rule{0mm}{2mm}
\end{table}

The Catalogue of the Milky Way Cluster Stars in the 2nd Galactic Quadrant (MWSCS.Q2) exists in machine-readable form only and can be retrieved from the CDS online archive.\footnote{ftp://cdsarc.u-strasbg.fr/pub/cats, http://vizier.u\mbox{-}strasbg.fr} The catalogue consists of 650 files with data on stars in areas with confirmed clusters. The information given for each star is shown in Table~\ref{tbl:stardat}.

\section{MWSCD.Q2: the Catalogue of Milky Way Cluster Data in the 2nd Galactic
Quadrant}\label{sec:appccat}

\begin{table}[ht]
\caption{Contents of the MWSCD main table}
\label{tbl:cludat}
\setlength{\tabcolsep}{2pt}
\begin{tabular}{rlcl}
\hline
Col& Label&Units&Explanations\\
\hline
\noalign{\smallskip}
 1 &No             &  ---   & MWSC number\\
 2 &Name           &  ---   & NGC, IC or other common designation\\
 3 &type           & ---    & object type\\
 4 &RA             &   h    & RA J2000.0 of the centre\\
 5 &DE             &  deg   & Dec J2000.0 of the centre\\
 6 &l              &  deg   & Galactic longitude of the centre\\
 7 &b              &  deg   & Galactic latitude of the centre\\
 8 &$r_0$          &  deg   & Angular radius of the core\\
 9 &$r_1$          &  deg   & Angular radius of the central part\\
10 &$r_2$          &  deg   & Angular radius of the cluster\\
11 &PMx            & mas/yr & average proper motion in\\
   &               &        & RA*cos(DE) \\
12 &PMy            & mas/yr & average proper motion in DE\\
13 &$\varepsilon_{PM}$            & mas/yr & Error of PM\\
14 &RV             &  km/s  & Average radial velocity\\
15 &$\varepsilon_{RV}$            &  km/s  & Error of RV\\
16 &nRV            &  ---   & Number of stars used for RV\\
17 &$N1\sigma_{r_0}$    & ---    & Number of most probable ($1\sigma$) members\\
   &               &        & in $r_0$\\
18 &$N1\sigma_{r_1}$    & ---    & Number of most probable ($1\sigma$) members\\
   &               &        & in $r_1$\\
19 &$N1\sigma_{r_2}$    & ---    & Number of most probable ($1\sigma$) members\\
   &               &        & in $r_2$\\
20 &d              & pc     & Distance from the Sun\\
21&$E(B-V)$        & mag    & Colour-excess in $(B-V)$\\
22 &$(K_s-M_{K_s})$& mag    & Apparent distance modulus\\
23 &$E(J-K_s)$     & mag    & Colour-excess in $(J-K_s)$\\
24 &$E(J-H)$       & mag    & Colour-excess in $(J-H)$\\
25 &$\Delta H$     & mag    & $\Delta H$\\
26 &$\log t$       & log yr & Logarithm of average age\\
27 &$\varepsilon_{\log t}$      & log yr & Error of $\log t$\\
28 &Nt             & ---    & Number of stars used \\
   &               &        & for the calculation of $\log t$\\
29 &$r_c$          & pc     & King' core radius\\
30 &$\varepsilon_{r_c}$& pc & Error of core radius\\
31 &$r_t$          & pc     & King' tidal radius\\
32 &$\varepsilon_{r_t}$& pc & Error of tidal radius\\
33 &$k$            & pc$^{-2}$& King' normalisation factor\\
34 &$\varepsilon_{k}$&pc$^{-2}$& Error of normalisation factor\\
35 &source         & ---    & Source for MWSC list and input\\ 
   &               &        & parameters\\
36 &source type    & ---    & Source object type\\
37 &[Fe/H]         & ---    & Metallicity\\
\noalign{\smallskip}
\hline
\end{tabular}\\
\end{table}

The Catalogue of the Milky Way Cluster Data in the 2nd Galactic Quadrant (MWSCD.Q2) exists in machine-readable form only and can be retrieved from the CDS online archive. The catalogue consists of the main table with the derived parameters of the confirmed clusters, the list of all clusters, and a notes file. The latter two files contain all surveyed clusters including those that were NOT confirmed. To inform the reader on the data scope included in the catalogue, we describe here the main table (Table~\ref{tbl:cludat}). Object types are indicated by A1 strings: empty - open cluster, a - association, g - globular cluster, m - moving group, n - nebulosity/presence of nebulosity, r - remnant cluster, $\ast$ - asterism. Source object types are given with A3 strings describing a larger variety of previous classifications, e.g. dubious clusters, clusters with variable extinction, embedded clusters, etc. The cluster metallicities have been adopted from the compilation of \citet{daml02}, their uncertainties are of about 0.1~dex.

\section{MWSCA.Q2: the Atlas of Milky Way Cluster Diagrams in the 2nd Galactic Quadrant}\label{sec:atl}

For each cluster we prepared two pages of diagrams that are combined in the atlas. For a better understanding, we describe the diagrams of the atlas taking the cluster King~19 as an example.

We consider the cluster map as the basic diagram of the first page, and the CMDs $K_s,(J-H)$ and $K_s,(J-K_s)$ as the basic diagrams of the second page. Stars are shown as coloured circles or dots. Symbols and their colours have the same meaning in all plots. Cyan symbols mark stars outside the cluster radius $r_2$, green symbols stars within $r_2$. The most probable kinematic and photometric members ($1\sigma$-members) are indicated in black for members located within $r_1$, red for members between $r_1$ and $r_2$, and blue for stars outside $r_2$. Cyan bars show the uncertainty for $1\sigma$-members (page 2).

\begin{figure*}[t]
\resizebox{\hsize}{!}{
\includegraphics[origin=l,angle=270,clip=]{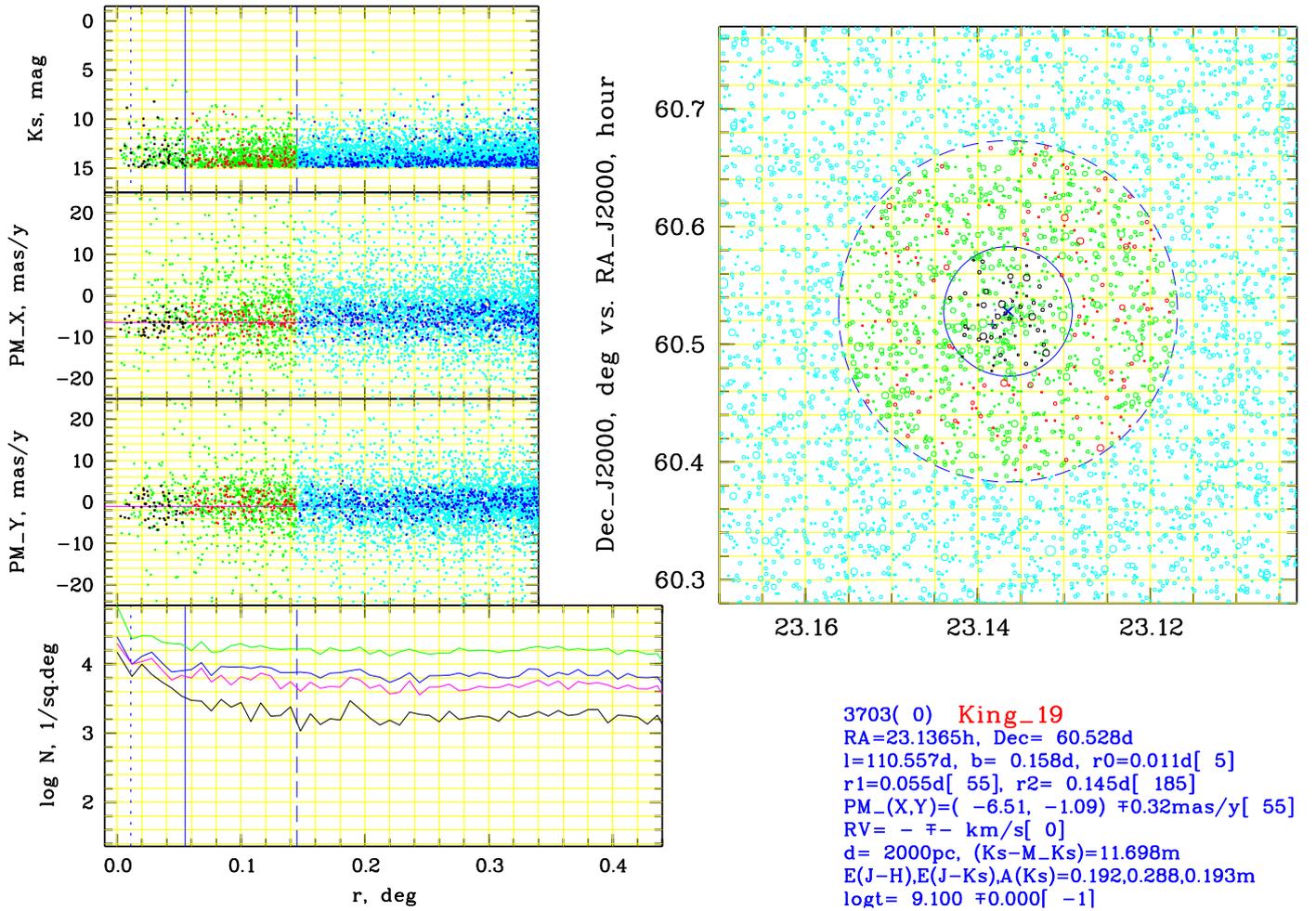}
}
\caption{Open cluster King~19 in the MWSC atlas (page~1). See text for the description.
}
\label{fig:atlp1}
\end{figure*}

Page~1 of the atlas (Fig.~\ref{fig:atlp1}) contains five diagrams with  spatial information and a legend on the derived cluster parameters. The right panel is a map of the cluster surrounding, the left panels show magnitudes $K_s$, proper motions $PM_x$, $PM_y$, and surface density $N$ versus distances $r$ of stars from the cluster centre.

The sky map: stars are shown by circles. The size corresponds to star brightness arranged in six $K_s$ magnitude bins. The blue cross indicates the cluster centre determined in this study, while the blue plus sign is the cluster position taken from the literature. If by chance other clusters appear in this cluster area, their centres are marked by magenta plus signs. Large blue circles (shown by dotted, solid, or dashed curves) indicate the cluster radii $r_0$, $r_1$, or $r_2$, respectively. The left panels: blue vertical lines (dotted, solid, or dashed) mark $r_0$, $r_1$, or $r_2$. Magenta horizontal lines in the $PM$ vs. $r$ diagrams correspond to the derived average proper motion of the cluster. The RDPs in the bottom panel are shown with green for all stars, blue for $3\sigma$-members, magenta for $2\sigma$-members and  black for $1\sigma$-members.

\begin{figure*}[t]
\resizebox{\hsize}{!}{
\includegraphics[origin=l,angle=270,clip=]{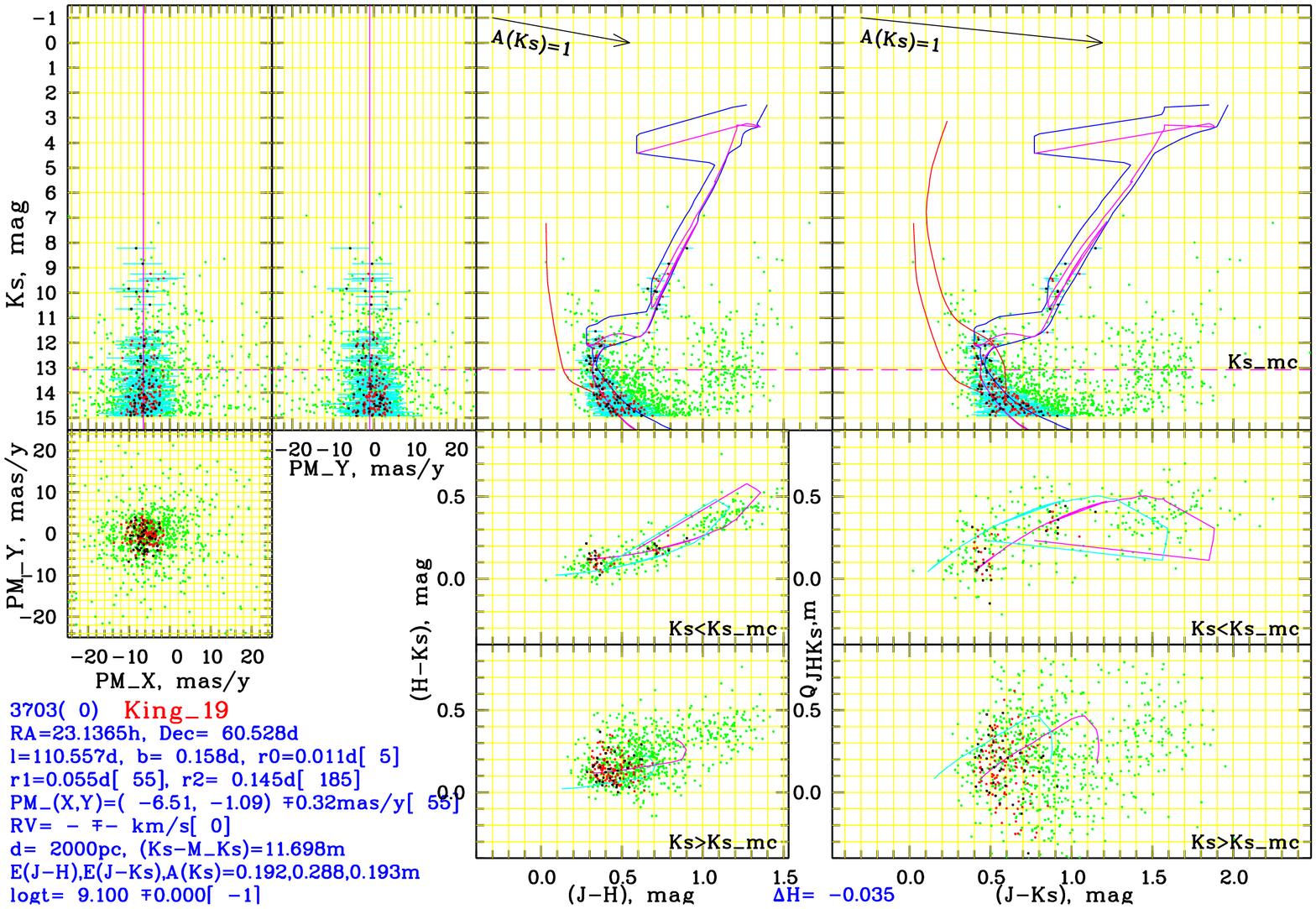}
}
\caption{Open cluster King~19 in the MWSC atlas (page~2). See text for the
description.
}
\label{fig:atlp2}
\end{figure*}

The legend gives cluster name, MWSC number, and COCD number in parentheses; equatorial $RA_{J2000}$, $Dec_{J2000}$, and galactic $l,\,b$ coordinates of the cluster centre; apparent cluster sizes $r_0$, $r_1$, $r_2$ and number of $1\sigma$-members within the corresponding radius; weighted average components $PM_{X,Y}$ of proper motion with their $rms$ errors and number of stars  used to compute the average; the average radial velocity, $RV$, $rms$ error, and the number of stars used to compute the average; distance to the cluster, $d$, distance modulus, $(K_s-M_{K_s})$; NIR interstellar reddening, $E(J-H),\,E(J-Ks)$, and interstellar extinction, $A(K_s)$; cluster age, its \textit{rms} error, the number in brackets gives the number of stars used to compute the average age, or it is -1 if an  isochrone fitting was applied. $\Delta H$ shown below the photometric diagrams indicates the empirical correction to the $H$-magnitude introduced in Sect.~\ref{sec:extinction} and \ref{sec:phopar}.

Page 2 (Fig.~\ref{fig:atlp2}) contains three diagrams with kinematic information (left panels), and six diagrams with photometric information (right panels).

The three left panels with kinematic data: the two upper diagrams show $PM_{X,Y}$ vs. $K_s$ relations, i.e.``PM-magnitude equation''. Magenta vertical lines correspond to the average proper motion of the cluster. The magenta dashed line shows the apparent magnitude $K_s^{mc}$, which corresponds to the bluest colour $(J-K_s)$ of the adopted isochrone. The bottom panel is the vector point diagram of proper motions.

The six right panels with photometric data: the two upper diagrams are CMDs ($K_s,(J-H)$ and $K_s,(J-K_s)$). The magenta curve is the apparent isochrone closest to the determined cluster age. Solid blue lines outline a domain of 100\% photometric members. Solid red lines are the ZAMS and TAMS (shown only in $K_s,(J-K_s)$). The magenta dashed line shows the apparent magnitude $K_s^{mc}$. The thick yellow circles mark the stars used for the age determination \citep[see][for details]{clucat}. The black arrows show the vectors of increasing extinction.   The four bottom panels show the two-colour $(H-K_s)/(J-H)$ diagram (left column) and $Q_{JHK}$-colour diagram (right column). The upper row is for stars brighter than $K_s^{mc}$, the lower row is for stars fainter than $K_s^{mc}$. Magenta curves indicate the apparent isochrone (i.e., observed colours), while cyan curves show the intrinsic isochrone.

The legend is the same as in page 1.

\end{document}